\documentclass{article}

\usepackage{arxiv}

\usepackage[utf8]{inputenc} 
\usepackage[T1]{fontenc}    
\usepackage{hyperref}       
\usepackage{url}            
\usepackage{booktabs}       
\usepackage{amsfonts}       
\usepackage{nicefrac}       
\usepackage{microtype}      
\usepackage{lipsum}
\usepackage{graphicx}
\usepackage[super]{nth}
\usepackage{xurl}

\title{Interactive SARS-CoV-2 mutation timemaps}

\author{
   Ren\'e L.~Warren\\
  Genome Sciences Centre, BC Cancer\\
  Vancouver, BC, V5Z 4S6, Canada \\
  \texttt{rwarren@bcgsc.ca} \\
   \And
 Inan\c{c} ~Birol \\
  Genome Sciences Centre, BC Cancer\\
  Vancouver, BC, V5Z 4S6, Canada \\
  \texttt{ibirol@bcgsc.ca} \\
}

\begin{document}
\maketitle

\begin{abstract}
As the year 2020 draws to an end, several new strains have been reported for the SARS-CoV-2 coronavirus, the agent responsible for the COVID-19 pandemic that has afflicted us all this past year. However, it is difficult to comprehend the scale, in sequence space, geographical location and time, at which SARS-CoV-2 mutates and evolves in its human hosts. To get an appreciation for the rapid evolution of the coronavirus, we built interactive scalable vector graphics maps that show daily nucleotide variations in genomes from the six most populated continents compared to that of the initial, ground-zero SARS-CoV-2 isolate sequenced at the beginning of the year.
\textbf{Availability:} Mutation time maps are available from \url{https://bcgsc.github.io/SARS2/}
\end{abstract}


\section{Introduction}
In the last few weeks of 2020, new severe acute respiratory syndrome coronavirus 2 (SARS-CoV-2) mutations in the United Kingdom (UK) have been reported \cite{rambaut}. Although coronavirus genome mutations have been previously discovered and announced throughout the year, including the widely discussed D614G missense change in the spike protein \cite{korber}, the latest recurring surface protein mutations to be identified (eg. N501Y, P681H) are cause for concern. The SARS-CoV-2 viral \textit{S} gene encodes a surface glycoprotein, which upon interaction with host ACE-2 receptors, makes it possible for the coronavirus to gain entry into host cells and propagate and the reported changes to its sequence may be associated with increased virulence \cite{gu}, infectivity \cite{korber} and overall fitness \cite{plante}. The global response to those recent reports has been swift, with several countries shutting down air travel from the UK. This highlights the severity of the situation and the importance to track genomic variations and their predicted effects over time and space.\par

The rapid evolution of the SARS-CoV-2 genome in human hosts has prompted us to map all nucleotide changes that have appeared in 2020, since the first genome sequence of a COVID-19 patient isolate from the outbreak epicentre in Wuhan, China was made public \cite{wu}. For this, we leveraged the collaborative efforts of hundreds of institutions worldwide who have graciously shared over 215,000 SARS-CoV-2 genome sequences with the GISAID central repository since early January 2020  \cite{gisaid}. Our mutation time maps show the staggering number of nucleotide variants that have accumulated on the whole viral genome throughout the year, and especially since fall 2020, and in the six most populated continents. Here we present key features of these maps and how they may be of utility to researchers.\par

\begin{figure}
    \centering
    \includegraphics[width=165mm]{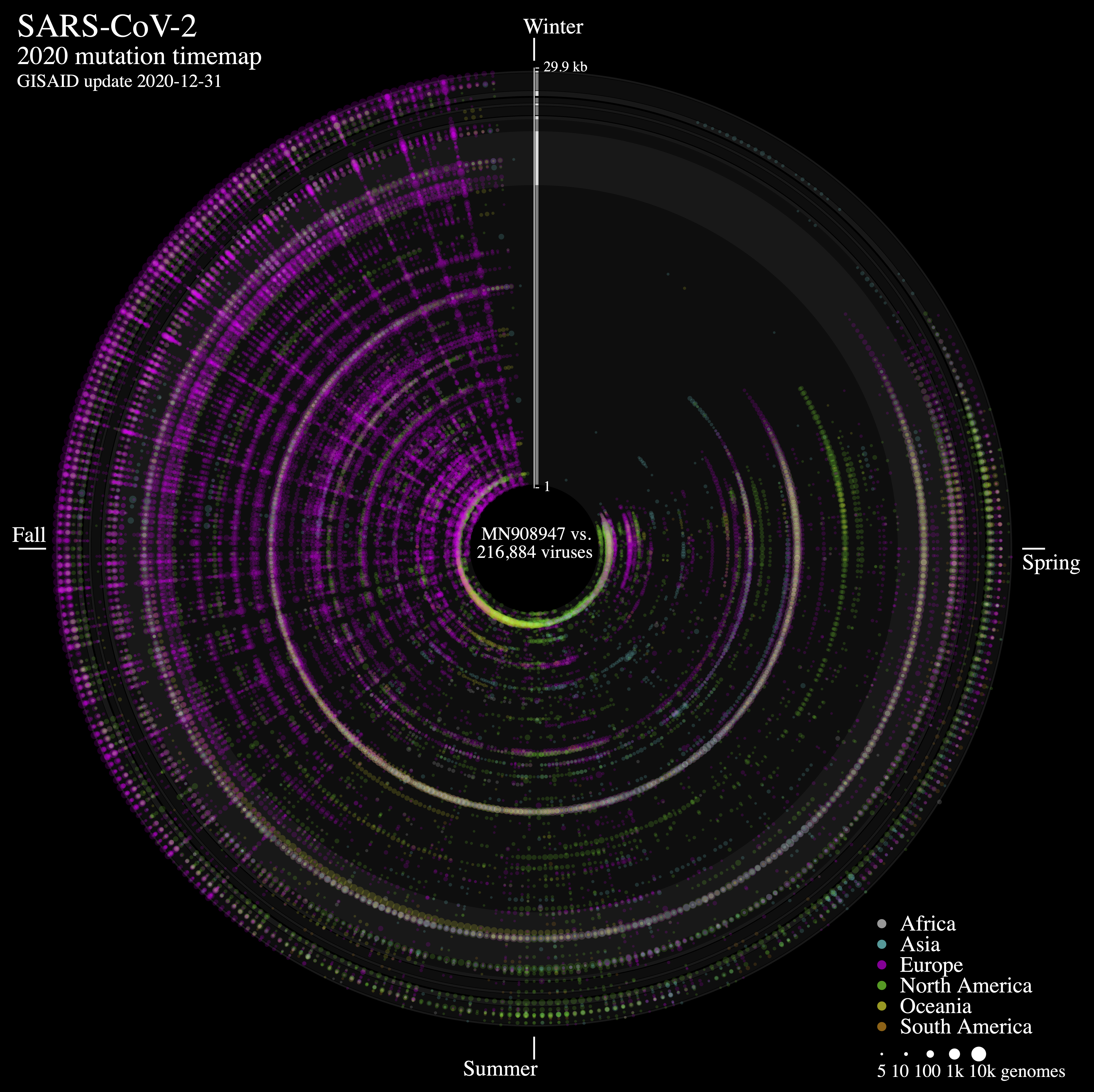}
    \caption{\textbf{SARS-CoV-2 evolution in human hosts.} ntEdit was used to map nucleotide variations between the first published coronavirus isolate from Wuhan, China in early January and over 215,000 SARS-CoV-2 genomes sampled from around the globe during the 2020 COVID-19 pandemic. This map shows missense mutations arising daily within the whole viral genome, with the reference genome represented by the vertical axis from bases 1 to 29.9 kbp. Alternating dark/light grey vertical rectangles and associated tracks depict, starting from the center, SARS-CoV-2 genes \textit{orf1ab, S, ORF3a, E, M, ORF6, ORF7a, ORF8, N,} and \textit{ORF10}. Mutations identified daily and throughout the viral genome are represented by circles in a given radius, and are coloured by continent of origin and sized relative to frequency occurring on the sample collection day. The 2020 calendar year mutations are organized clockwise from the upper vertical. Hovering the mouse pointer/cursor over a mutation (not shown) reveals the day, nucleotide change, continent of origin with frequency occurring and predicted amino acid change, when applicable. The latter feature is useful to quickly identify missense variants with the potential to alter protein function. The sparse mutation signature observed in late December is due to a lag between genome collection and submission to GISAID. Additional maps are available from: \url{https://bcgsc.github.io/SARS2/}.}
  \label{fig:fig1}
\end{figure}

\section{Methods}
We first downloaded all complete, high-coverage SARS-CoV-2 genomes from GISAID \cite{gisaid} (human hosts samples collected in 2020; \url{https://www.epicov.org/}). We then ran a genome polishing pipeline, which consists of ntHits \cite{mohamadi} (v0.1.0 -b 36 --outbloom -c 1 -p seq -k 25) followed by ntEdit \cite{warren} (v1.3.4 -s 1 -r seq\_k25.bf). We used the first published SARS-CoV-2 genome isolate \cite{wu} (WH-Human 1 coronavirus, GenBank accession: MN908947.3) as the reference and each individual GISAID genome in turn as source of kmers to identify base variation relative to the former. The variant call format (VCF) output files from ntEdit were parsed and we tallied, for each submitted GISAID genome, the complete list of nucleotide variations. We next organized each nucleotide variant by sample collection date, continent of origin and, when applicable, evaluated its effect on the gene product that harbours the change to output an interactive scalable vector graphics (SVG) file.\par

\section{Results and Discussion}
We analyzed nucleotide variations over time in more than 215,000 SARS-CoV-2 viral genomes, submitted to the GISAID initiative \cite{gisaid} from around the globe, relative to that of the ground zero COVID-19 clinical isolate \cite{wu}. We mapped each mutation that was observed in five or more genomes each day. The 2020 calendar year from January \nth{1} 2020 (day 1) to December \nth{24} 2020 (day 359) is organized in a circle where each radius represents a day and data points represent mutations along the reference genome sequence from 1 (closest to center) to 29,903 bp (near the outer rim). The size of each point is in log10 scale of the number of contributing viral genomes collected on that day that has the mutation, with colour assignments indicating the continent of origin where the mutation is observed. A mouse over each data point reveals the collection date, the nucleotide variant, the continent and associated number of contributing genome sequences and, when applicable, the gene product and predicted amino acid change.

From the SARS-CoV-2 genome mutation time map (Fig. 1), we observe the first persistent mutations ($\geq$5 genomes/day) appearing in late February 2020, including the prevalent D614G mutation in Europe on February \nth{22} (albeit since January in fewer samples, not shown). From there, the original coronavirus genome sustained many changes overtime (4,674 distinct variants mapped as of December \nth{31} 2020), including a sizeable proportion (57.0 \%) of missense mutations. It is immediately evident from Fig. 1 that variations from Europe account for a larger share (72.5\%) of the variants mapped. Further, there appears to be a surge in variations identified in late summer/throughout fall 2020 in this continent. This may be explained by a disproportionate number of submissions with samples originating from this jurisdiction as the second wave hit hard. Thus, caution in interpreting the map is warranted. Of note, the observation of the spike protein gene variant N501Y observed on our maps in Europe in late September 2020 and consistent with an earlier study reporting on its recurrent emergence within this time frame \cite{rambaut}. We think these maps will be of utility to researchers in their exploration of SARS-CoV-2 mutations and their predicted effect over time.\par

\subsection*{Grant information}
This work was supported by Genome BC and Genome Canada [281ANV]; and the National Institutes of Health [2R01HG007182-04A1]. The content of this paper is solely the responsibility of the authors, and does not necessarily represent the official views of the National Institutes of Health or other funding organizations.

\subsection*{Acknowledgements}
We acknowledge Cecilia Yang for her early work on SARS-CoV-2 variants.

\bibliographystyle{unsrt}  


%

\end{document}